\documentclass[conference]{IEEEtran}
\IEEEoverridecommandlockouts
\usepackage{cite}
\usepackage{amsmath,amssymb,amsfonts}
\usepackage{algorithmic}
\usepackage{graphicx}
\usepackage{textcomp}
\usepackage{xcolor}
\usepackage{float}
\usepackage[caption=false, font=footnotesize]{subfig}
\usepackage{enumitem,amssymb}
\newlist{todolist}{itemize}{2}
\setlist[todolist]{label=$\square$}

\def\BibTeX{{\rm B\kern-.05em{\sc i\kern-.025em b}\kern-.08em
    T\kern-.1667em\lower.7ex\hbox{E}\kern-.125emX}}
\begin{document}

\title{Analysis of Processing Pipelines for Indoor Human Tracking using FMCW radar
}

\author{\IEEEauthorblockN{Dingyang Wang, Francesco Fioranelli and Alexander Yarovoy}
\IEEEauthorblockA{\textit{Microwave Sensing Signal \& Systems (MS3) Group, Department of Microelectronics} \\
\textit{Delft University of Technology, Delft, The Netherlands }\\
\{D.Wang-6, F.Fioranelli, A.Yarovoy\}@tudelft.nl}
}

\maketitle

\begin{abstract}
In this paper, the problem of formulating effective processing pipelines for indoor human tracking is investigated, with the usage of a Multiple Input Multiple Output (MIMO) Frequency Modulated Continuous Wave (FMCW) radar. 
Specifically, two processing pipelines starting with detections on the Range-Azimuth (RA) maps and the Range-Doppler (RD) maps are formulated and compared, together with subsequent clustering and tracking algorithms and their relevant parameters. Experimental results are presented to validate and assess both pipelines, using a 24 GHz commercial radar platform with 250 MHz bandwidth and 15 virtual channels. Scenarios where 1 and 2 people move in an indoor environment are considered, and the influence of the number of virtual channels and detectors' parameters is discussed. The characteristics and limitations of both pipelines are presented, with the approach based on detections on RA maps showing in general more robust results.
\end{abstract}


\section{Introduction}

Accurate human positioning \& tracking in indoor environments is an interesting problem in many applications and environments such as airports, public spaces, shopping malls. 

Video cameras for human tracking \cite{ObjectTrackingSurveyyilmaz2006} have been researched for a long time. With the aid of a stereoscopic sensor, the distance can be measured. However, privacy concerns are high with this approach. On the other hand, the Lidar sensor is another approach widely used in automotive research for mapping and sensing environments. But, in indoor environments, Lidar is sensitive to ambient light and might raise issues of lasers being used in confined spaces. 
As a sensor capable of estimating the range and velocity information of targets, radar has attracted considerable interest for indoor tracking, also considering its (relative) safety for privacy as no videos or pictures are recorded. With the progress in electronic and manufacturing technologies, radar sensors are becoming more compact and portable. Furthermore, the development of Multiple Input Multiple Output (MIMO) radar supported by the automotive domain delivered the capability to estimate the angular position of targets, in both azimuth \& elevation, besides the more conventional estimations of range and velocity. 

This makes radar a very candidate for indoor human tracking \cite{MultipleTargetPositioningyoo2019}, with different approaches presented in the literature, either based on the Range-Azimuth (RA) domain \cite{PeopleTrackingCountingtexasinstruments}, or the Range-Doppler (RD) domain \cite{RadarBasedRobustPeopleninos2022,NoncontactExtractionBiomechanicalwang2021,GroupedPeopleCountingren2023}. It should be noted that the performances that these approaches can achieve are strongly dependent on the specification of the radar and of the processing pipelines' parameters. In this paper, two processing pipelines starting from detections on the RA and RD maps are formulated and compared, assessing the effect of different operational parameters, such as detectors' parameters and the number of azimuth channels determining the angular resolution. Experiments are performed using a 24 GHz MIMO radar for validation.  

The paper is organized as follows. In Section \ref{sec:pipe}, the general signal processing steps for indoor human tracking are reviewed and the two proposed pipelines are described. In Section \ref{sec:ExpResults}, the experimental setup is described and performance analysis is provided. Specifically, various types of Constant False Alarm Rate (CFAR) detectors are compared and the impact of the number of azimuth channels is assessed, for scenarios with 1 and 2 people moving in the indoor environment. Finally, conclusions are presented in Section \ref{sec:conclusion}.

\section{Proposed Processing Pipelines}
\label{sec:pipe}

\begin{figure}[t]
\centering
\subfloat[Detection on RA map]{\label{fig:RAdiagram}\includegraphics[width=0.35\linewidth]{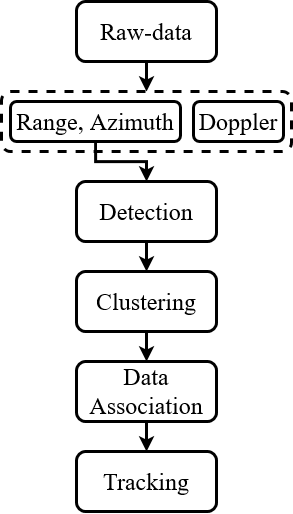}}
\,
\subfloat[Detection on RD map]{\label{fig:RDdiagram}\includegraphics[width=0.35\linewidth]{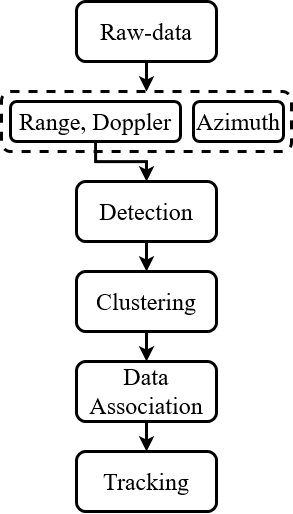}}
\caption{Block diagram of the proposed processing pipelines.}
    \label{fig:pipeline}
\end{figure}

The Frequency Modulated Continuous Wave (FMCW) radar signal transmitted in the time domain can be defined as \cite{JointDFTESPRITEstimationkim2015}:
\begin{equation}
\label{eq:signalModel}
x_0(t)=\exp \left(j2\pi \left(f_ct+\frac{\mu}{2}t^2 \right)\right)
\end{equation}
where $f_c$ is the carrier frequency, $\mu$ is the rate of change (slope) of the instantaneous frequency of the chirp signal (i.e., $\mu = f_{BW}/T_c$), $f_{BW}$ is the bandwidth, and $T_c$ is the sweep time.
The range resolution is related to the bandwidth by $\Delta r = c/2f_{BW}$, where $c$ is the speed of light. By transmitting multiple chirps in a frame to increase the Doppler resolution, its value is defined as $\Delta v =\lambda / (2 N_{c}T_c)$,
where $\lambda$ is the wavelength, and $N_c$ is the number of chirps in a frame. Furthermore, using an antenna array with $N_{an}$ elements enables the estimation of the direction of arrival (DOA) from the targets. The angular resolution is given by $\Delta \theta = \lambda /(N_{an} d cos(\theta))$, where $d$ is the antenna spacing, and $\theta$ is the angle in the radial direction. When $\theta =0$ at boresight, the best angular resolution is obtained.
To track multiple people using a radar, advanced signal processing algorithms have been formulated and applied in the literature \cite{RealTimePeopleTrackingpegoraro2021,GroupedPeopleCountingren2023,RadarBasedRobustPeopleninos2022,LowComplexityRadarDetectorsafa2021}.
Inspired by these works, in this paper two processing pipelines are formulated and implemented, one based on RA maps and one on RD maps, as shown in Fig. \ref{fig:pipeline}. In general, their main processing steps 
are summarized as follows:
\begin{enumerate}
    \item \textbf{Pre-processing}: The typical pre-processing steps of MIMO FMCW radar involve the de-chirping and the application of Fast Fourier Transform (FFT) to infer range information from the de-chirped signal. Furthermore, a second FFT is often applied across a given number of chirps (i.e., coherent processing interval CPI) to extract Doppler information. Finally, the angle information can also be extracted based on the phase differences between different channels, often using an FFT-based beamforming or more advanced techniques. The final product is the so-called 'radar cube' containing range, Doppler, and angular information. Additionally, part of the pre-processing steps include clutter filtering and background subtraction that can help enhance the detection performance.
    \item \textbf{Detection}: A detector is an algorithm used for extracting relevant targets' information against the background noise and clutter components. Constant False Alarm Rate (CFAR) detectors are a widely used signal processing technique in radar systems to detect targets in the presence of background clutter and noise, while maintaining a constant false alarm rate \cite{LowComplexityRadarDetectorsafa2021}. The primary goal of CFAR-based detection is to set an appropriate detection threshold that can also adapt to changes in the background environment, ensuring reliable target detection in various operating conditions. With the development of these algorithms, their implementation can be applied to 1D range profiles, or to 2D maps such as RA maps or RD maps. 
    In section \ref{sub:CFAR}, the performance of different 2D-CFAR detectors is  discussed. 

    \item \textbf{Clustering}: Human targets observed by millimeter-wave radar appear as "extended targets" due to the short wavelength relative to the size of the body and the movement from different parts of the human body resulting in Doppler dispersion. This leads to the presence of multiple detected cells in the RA or RD maps for a single target, which clustering algorithms can process. A commonly used clustering method for mm-wave radars and extended targets is the Density-Based Spatial Clustering of Applications with Noise (DBSCAN) algorithm \cite{DensitybasedAlgorithmDiscoveringester1996}. DBSCAN is particularly well-suited for radar signal processing due to its ability to handle irregularly shaped clusters and automatically identify outlier points within the data. This step can help differentiate between multiple targets and reduce false alarms, even if its performance can be reduced when multiple targets are close to each other, within one resolution cell.
    
    \item \textbf{Data Association}: Tracking algorithms are used to maintain estimates of each target's position, velocity, and other relevant parameters over multiple observations over time. Data association techniques are applied to assign new/current radar detections to exist tracks, and when needed establish new tracks for newly detected targets based on distance metrics (e.g., the Mahalanobis distance in Global Nearest Neighbor - GNN) or probability (e.g., Joint Probability Data Association - JPDA \cite{ProbabilisticDataAssociationyaakov2009}). In this paper, the probability-based method JPDA was used for data association. 

    \item \textbf{Tracking}: In this paper, an Extended Kalman filter (EKF) was used as a tracking algorithm to predict and update the state according to the chosen kinematic model. The EKF performs an analytic linearization operation, due to the conversion from polar to Cartesian coordinates. The model also includes the range, angle, and radial velocity $(r,\theta, \dot r)$ components from the measurements. In addition, the algorithm also defines history-based track management for the generation of new tracks and the deletion of no updated tracks.

\end{enumerate}

Besides the general steps discussed above, additional details on the detection steps based on RA and RD maps as shown in Fig. \ref{fig:pipeline} are presented in the following sub-sections.


\subsection{Detection on RA maps}
As radar systems are physically designed with an increasing number of channels, their angle resolution is also increasingly improved. The detection on RA maps can therefore be an approach that extracts representative point clouds for short-range indoor people tracking. For instance, one approach proposed by Texas Instruments \cite{PeopleTrackingCountingtexasinstruments} shows to localize people in 3D coordinates up to 15 m range and a counting density of 1 person per square meter. Another example is derived from marine target localization \cite{24GHzFMCWgennarelli2022}, but with the common aspect that the targets in the context of this work and with the considered radar, are also distributed/extended. 

In this paper, the proposed algorithm pipeline of detection on RA map $I_{RA}(r, \theta)$ is shown in Fig. \ref{fig:RAdiagram}, where $r$ denotes the range and $\theta$ denotes the (azimuth) angle information. After range-angle FFT, a 2D CFAR detector is applied on the resulting RA map, which produces a binary image $I_{RAd}(r, \theta)$. Before clustering the points from the binary image $I_{RAd}(r, \theta)$, a conversion from polar coordinate to Cartesian coordinate is needed. After integration with the Doppler information for each detected cell, the point clouds can be represented by a sequence of triplets $(r,\theta, \dot r)$, where $\dot r$ is the radial velocity derived from the Doppler measurement. 
Afterward, the positions of detected objects are considered as measurements given in input to the data association and tracking algorithms, which estimate the position and velocity of targets and help reduce the number of false alarms. However, this processing pipeline may have limitations when considering multiple targets, especially when they are getting too close to each other and the spacing is less than the angular resolution. This is particularly noticeable at a farther distance from the radar or away from the boresight to the sides of the radar field of view and will cause some missed detections to be dealt with.


\subsection{Detection on RD maps}
In \cite{RadarBasedRobustPeopleninos2022,NoncontactExtractionBiomechanicalwang2021,GroupedPeopleCountingren2023}, an alternative signal processing pipeline is presented where the detector is applied on RD maps $I_{RD}(r, \dot r)$. In modern mm-wave radars, range resolution and Doppler resolution are generally good enough to separate objects, as they are located in different range bins or Doppler bins, or both. However, in the context of indoor human tracking, some cases can happen when two or more people move close to each other and walk next to each other, shoulder by shoulder or one behind the other. In these cases of 'grouped people', the conventional range and Doppler resolution may be insufficient for reliable discrimination and tracking of all individuals, and addressing this remains an open problem. 
In the processing pipeline, after the CFAR detection the objects are represented by 1 in binary images $I_{RDd}(r, \dot r)$. Then, the detected points are associated with their corresponding angle information via azimuth DOA estimation. Before clustering the resulting point clouds, the polar coordinates need to be converted into Cartesian coordinates to fit the distance criterion in the clustering algorithm. The remaining steps are then the same as for the detection on RA maps discussed in the previous subsection.

\section{Experimental results comparison}
\label{sec:ExpResults}

\begin{figure}[h]
\centering
\subfloat[Experimental room]{\label{fig:expEnv1}
\includegraphics[width=0.45\linewidth]{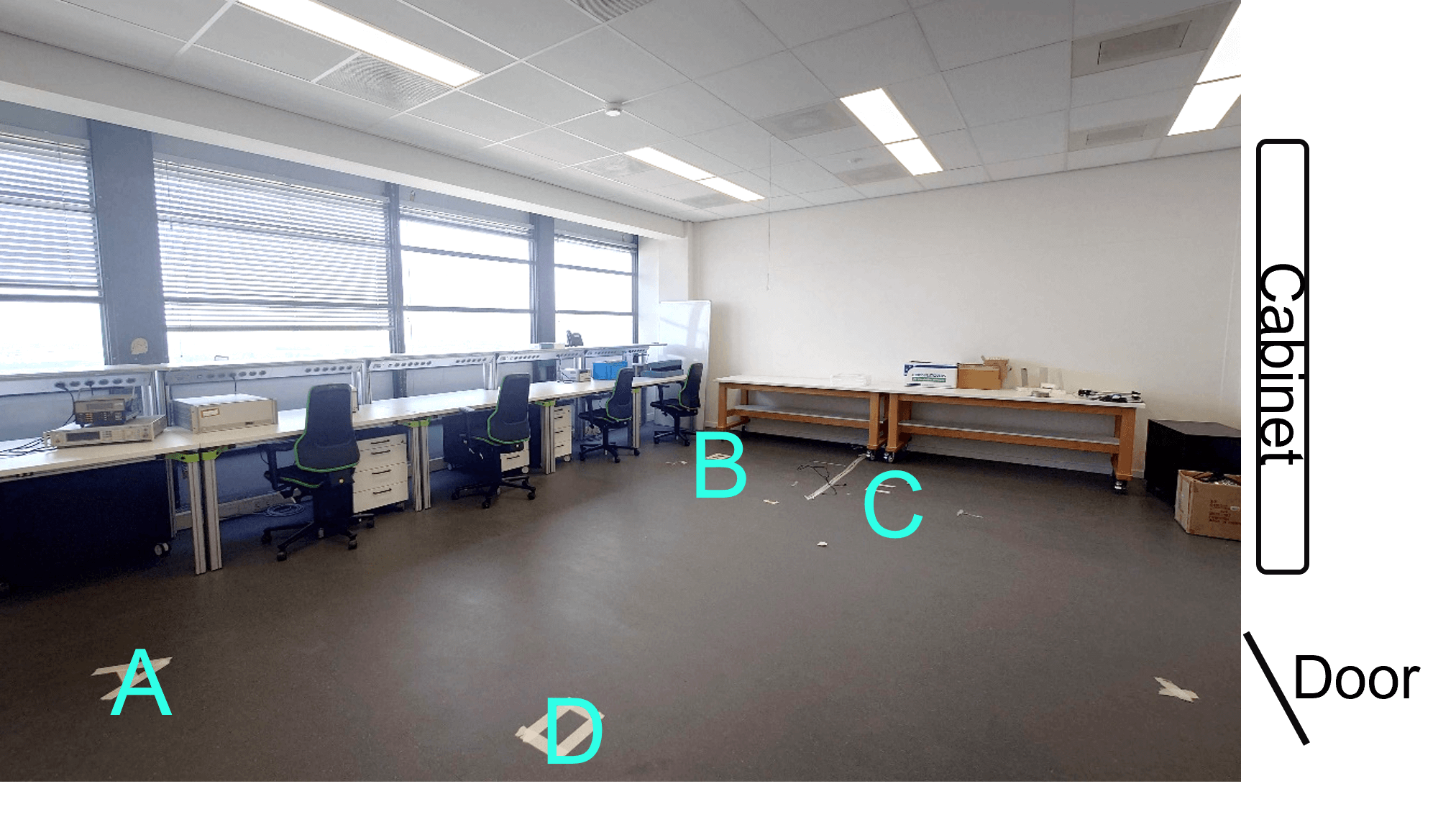}}
\,
\subfloat[Setup of radar and laptop]{\label{fig:expEnv2}
\includegraphics[width=0.45\linewidth]{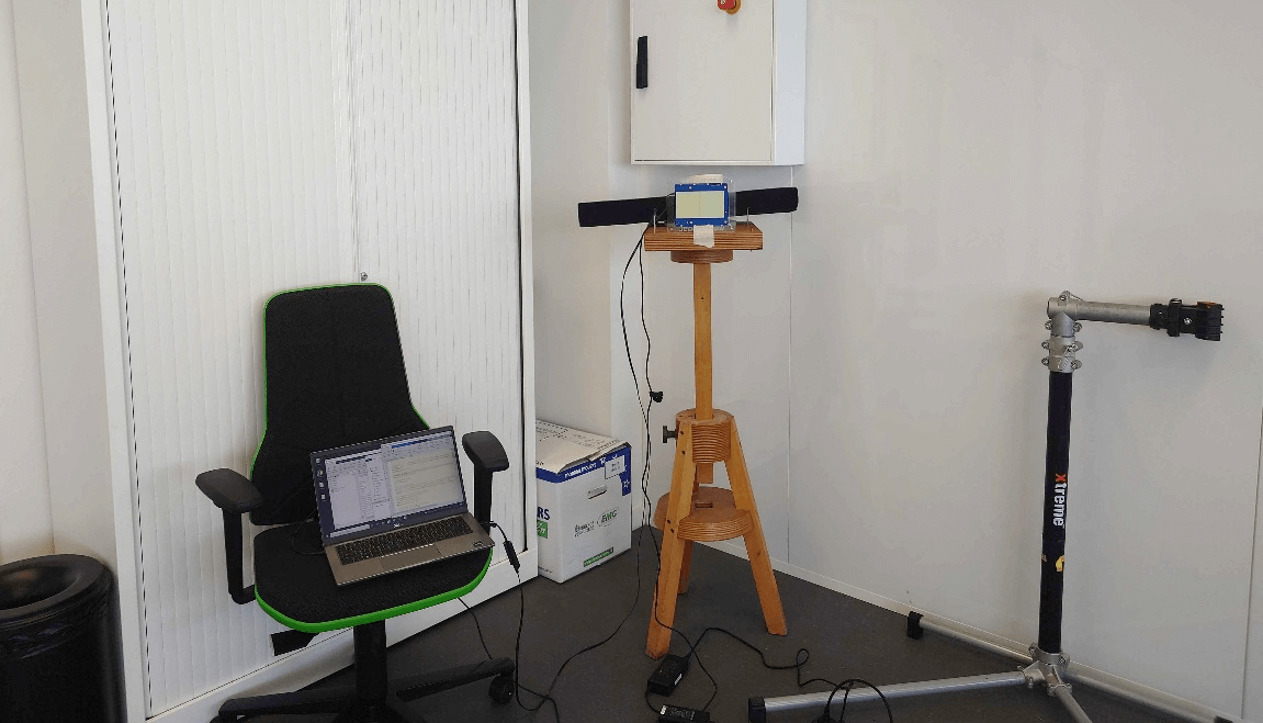}}
\caption{Experimental environment for data collection in the MS3 Radar Laboratory at TU Delft}
\label{fig:expEnv}
\end{figure}

For the purpose of this research, a data set was specifically collected with 5 individuals and 10 activities in the laboratory room of the MS3 group at TU Delft. To emulate the cluttered environment of normal office space, pieces of furniture such as tables, chairs, and cabinets were placed in the environment, and a metallic curtain was also present at the window, which contributed to the multi-path. As illustrated in Fig. \ref{fig:expEnv1}, the marked ABCD points are reference points to guide the movements of the participants along different trajectories. 
The radar was placed at around 1.3-meter height in the corner of the room, with its line of sight pointing along the diagonal of the room to get wide coverage, as shown in Fig. \ref{fig:expEnv2}. 

In this paper, a commercial 24GHz FMCW radar (by Joby Austria, former INRAS) with a relatively low bandwidth of 250MHz is used to evaluate the performance. The relatively small bandwidth results in a range resolution of approximately 60 cm, which indicates that the target occupies a large area in the range profile while also making it difficult to isolate individual body parts. The detailed parameters used are listed in Table \ref{tab:fmcwpara}.


\begin{table}[h]
\centering
\caption{JOBY (former INRAS) FMCW radar parameters}
\label{tab:fmcwpara}
\begin{tabular}{ll}
\hline
FMCW radar model                        &  RadarBook2 (RBK2)     \\
Operating frequency                     & 24 GHz         \\
Sweep bandwidth                         & 250 MHz        \\ 
ADC sampling rate                       & 120 ksps        \\
ADC samples                             & 56              \\
Up chirp duration                       & 467 $\mu$s          \\
Chirp repetition interval               & 483 $\mu$s          \\
Number of chirps in a frame             & 90 \\
Slow-time sampling frequency            & 10 Hz         \\
Number of TX \& RX channels             & 2 x 8 \\
Antenna horizontal 3 dB beamwidth       & 76.5$^{\circ}$ \\
\hline
\end{tabular}
\end{table}

\subsection{Analysis of CFAR detectors on RA and RD maps} 
\label{sub:CFAR}

Representative CFAR methods considered are cell-averaging (CA) \cite{PrinciplesModernRadarrichards2010}, greatest-of (GO) \cite{AnalysisModifiedCellAveragingweiss1982}, smallest-of (SO) \cite{RangeResolutionTargetstrunk1978} and order statistics (OS) \cite{NewCFARprocessorBasedhermannrohling1985} CFAR.
The process of detecting the presence of a target in the cell under test begins with comparing the radar measurement with a calculated threshold. The calculation of the threshold changes for each specific CFAR type, but depends on the desired probability of false alarm $P_{FA}$ and on the number of the training cells used to estimate the background level, and the guard cells used to prevent energy leakage in a cell under test.
The OS-CFAR is designed to suppress target masking in a multi-target environment and widely used in previous research with mm-wave radars \cite{PeopleTrackingCountingtexasinstruments, AngleInsensitiveHumanMotionzhao2022}. However, when operating on 2D data rather than on range or Doppler profiles, due to the relatively high computational complexity of sorting the cells' values, it is hard to achieve real-time performance. Therefore, a modified version of OSCA-CFAR was proposed by \cite{FastTwoDimensionalCFARkronauge2013}. 

\begin{figure*}[t]
\centering
\subfloat[CA-CFAR]{\label{fig:RA_CA_all}
\includegraphics[width=0.30\linewidth]{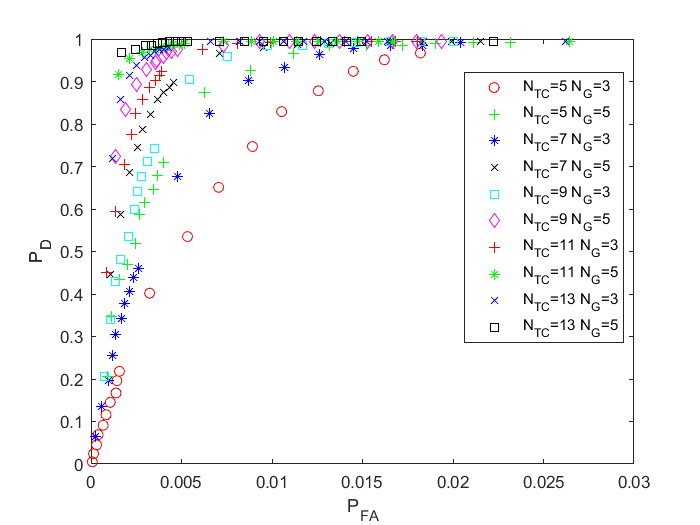}}
\,
\subfloat[OS-CFAR]{\label{fig:RA_OS_all}
\includegraphics[width=0.30\linewidth]{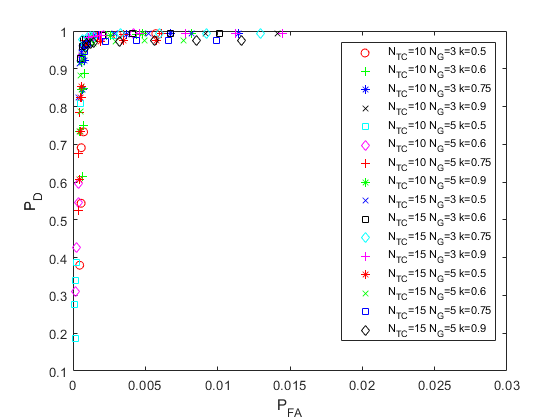}}
\,
\subfloat[OSCA-CFAR]{\label{fig:RA_OSCA_all}
\includegraphics[width=0.30\linewidth]{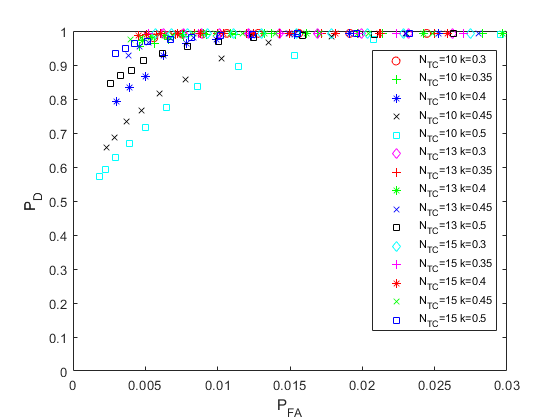}}

\caption{ROC curves for different combinations of $N_{TC}$ and $N_G$ parameters for different CFAR detectors; pipeline with detection on RA maps}
\label{fig:cfarRAmapresults}
\end{figure*}

\begin{figure*}[t]
\centering
\subfloat[CA-CFAR]{\label{fig:RD_CA_all}
\includegraphics[width=0.3\linewidth]{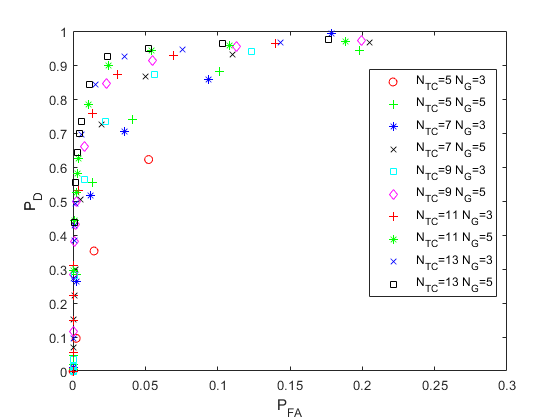}}
\,
\subfloat[OS-CFAR]{\label{fig:RD_OS_all}
\includegraphics[width=0.3\linewidth]{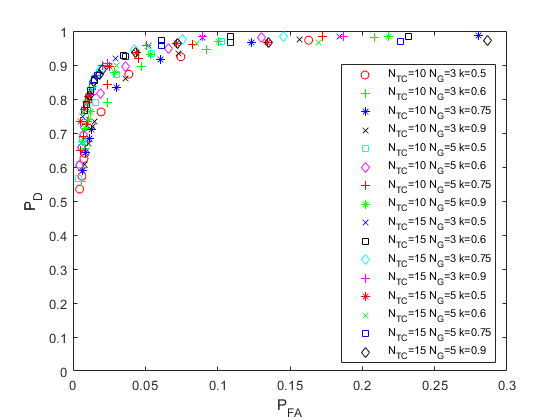}}
\,
\subfloat[OSCA-CFAR]{\label{fig:RD_OSCA_all}
\includegraphics[width=0.3\linewidth]{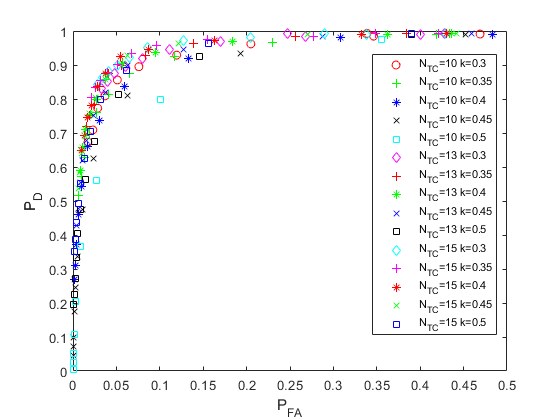}}

\caption{ROC curves for different combinations of $N_{TC}$ and $N_G$ parameters for different CFAR detectors; pipeline with detection on RD maps }
\label{fig:cfarRDmapresults}
\end{figure*}

In order to compare the performance of different CFAR detectors, the aforementioned five methods have been implemented on RA and RD maps containing one person moving in an indoor scenario. The receiver operating characteristic (ROC) curve is used to quantify the performance as a function of different parameters such as the number of training cells $N_{TC}$ and the number of guard cells $N_G$, as shown in Fig. \ref{fig:cfarRAmapresults} and Fig. \ref{fig:cfarRDmapresults}. Only the results for CA-CFAR, OS-CFAR and OSCA-CFAR are reported here for conciseness. It can be seen that in general, the OS-CFAR has the highest probability of detection $P_D$ for a given $P_{FA}$.
Comparing the two pipelines, the detection on RD maps produces more false alarms than the detection on RA maps, i.e., the lower area under the ROC curve for the same detector and parameters. Such false alarms can come from the indoor environment, such as multipath reflections from the walls, metallic curtains and furniture.

\subsection{Analysis as a function of number of channels}\label{Ana:channels}
\begin{figure*}[!t]
\centering
\subfloat[15 Channels]{\label{fig:FB_1T_15RX_RA}
\includegraphics[width=0.19\linewidth]{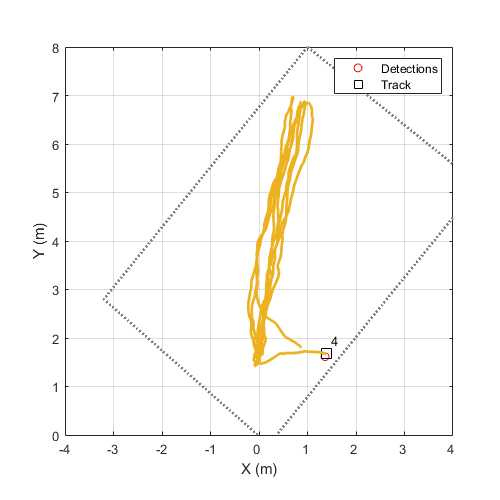}}
\subfloat[12 Channels]{\label{fig:FB_1T_12RX_RA}
\includegraphics[width=0.19\linewidth]{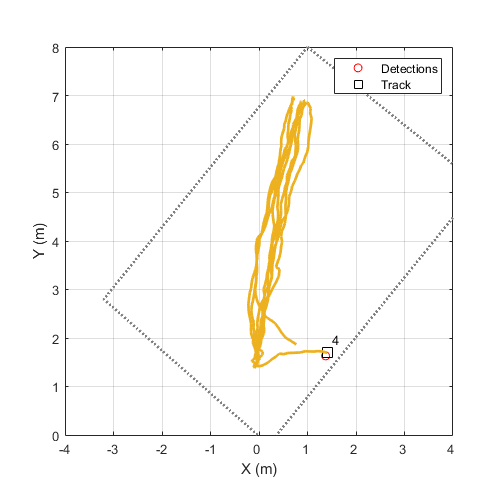}} 
\subfloat[8 Channels]{\label{fig:FB_1T_8RX_RA}
\includegraphics[width=0.19\linewidth]{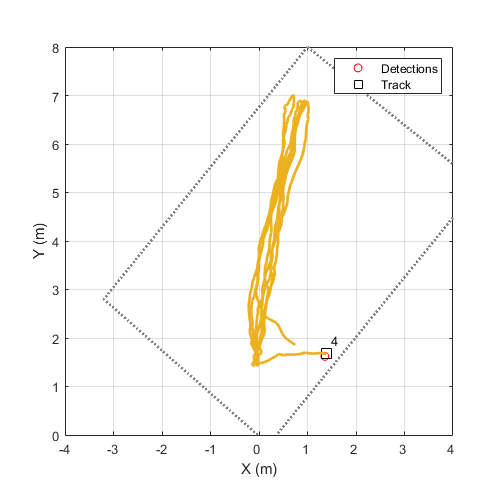}}
\subfloat[6 Channels]{\label{fig:FB_1T_6RX_RA}
\includegraphics[width=0.19\linewidth]{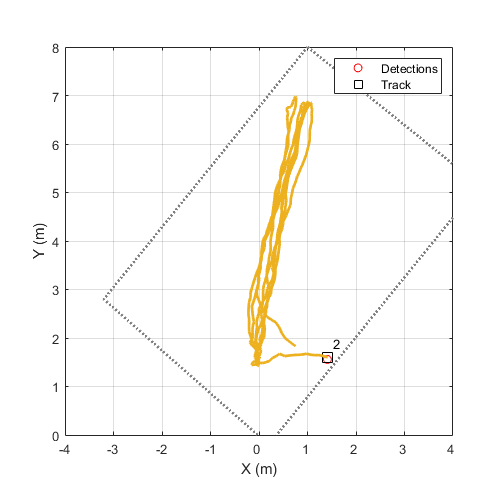}} 
\subfloat[4 Channels]{\label{fig:FB_1T_4RX_RA}
\includegraphics[width=0.19\linewidth]{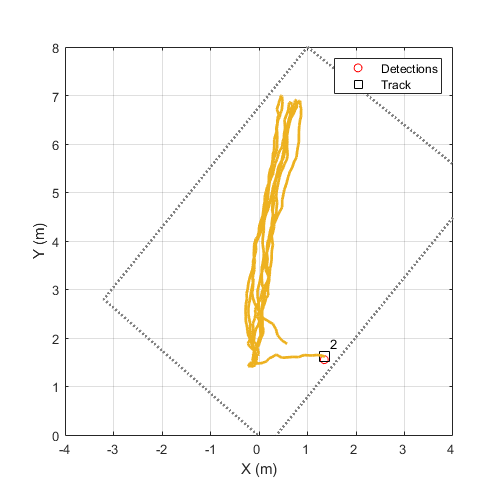}}
\caption{Number of channels (i.e., angular resolution) analysis for 1 person; pipeline with detection on RA maps}
\label{fig:re1TFBRA_CHN}
\end{figure*}

\begin{figure*}[!t]
\centering
\subfloat[15 Channels]{\label{fig:FB_1T_15RX_RD}
\includegraphics[width=0.19\linewidth]{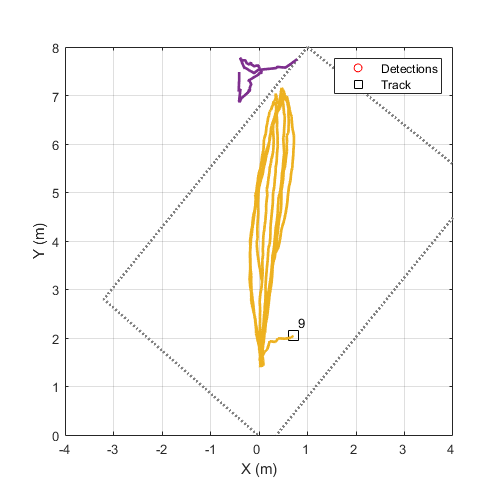}}
\subfloat[12 Channels]{\label{fig:FB_1T_12RX_RD}
\includegraphics[width=0.19\linewidth]{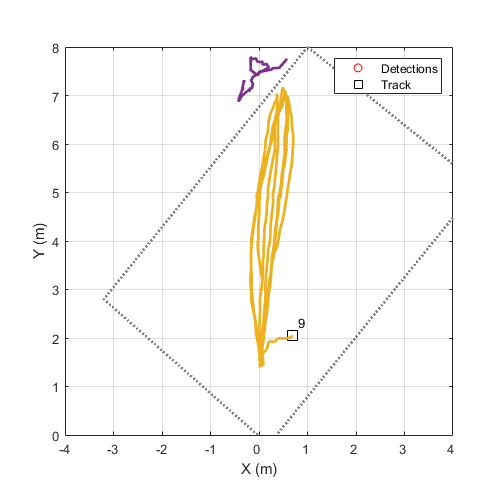}} 
\subfloat[8 Channels]{\label{fig:FB_1T_8RX_RD}
\includegraphics[width=0.19\linewidth]{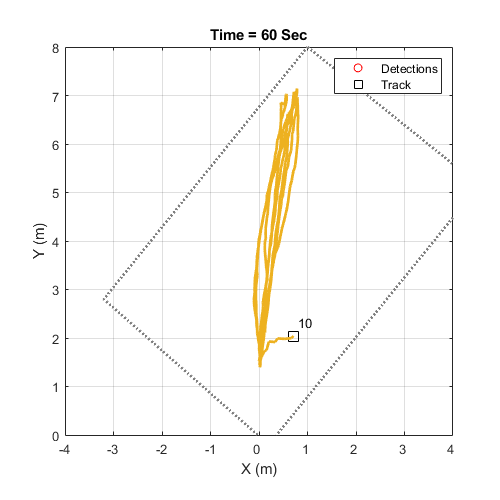}}
\subfloat[6 Channels]{\label{fig:FB_1T_6RX_RD}
\includegraphics[width=0.19\linewidth]{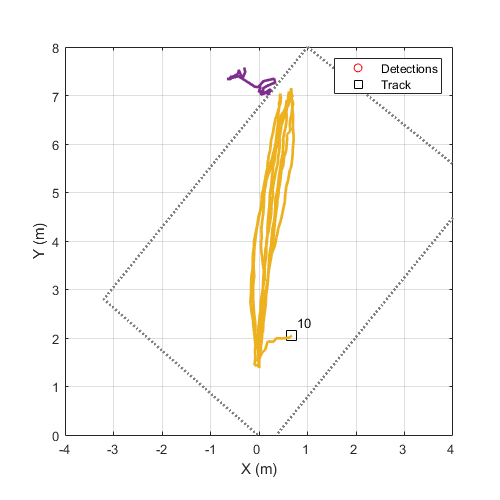}} 
\subfloat[4 Channels]{\label{fig:FB_1T_4RX_RD}
\includegraphics[width=0.19\linewidth]{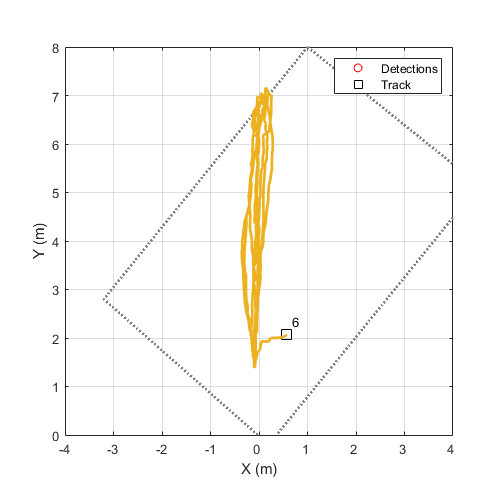}}
\caption{Number of channels (i.e., angular resolution) analysis for 1 person; pipeline with detection from RD maps}
\label{fig:re1TFBRD_CHN}
\end{figure*}

Experimental data are collected with the complete 15 virtual channels available in the RBK2 radar. Then, the effect of the number of channels for subsequent processing is evaluated in two scenarios.
Firstly, in the scenario with one single target shown in Fig. \ref{fig:re1TFBRA_CHN} and Fig. \ref{fig:re1TFBRD_CHN}, it appears to be no significant changes in the tracked trajectories when reducing the number of channels used in angle estimation for both pipelines on RA and RD maps.
In Fig. \ref{fig:re1TFBRD_CHN}e, the trajectories appear to shift from the marker point B when using 4 channels with the RD pipeline, whereas they remain the same for the RA pipeline.

\begin{figure*}[!t]
\centering
\subfloat[15 Channels]{\label{fig:FB_2T_15RX_RA}
\includegraphics[width=0.19\linewidth]{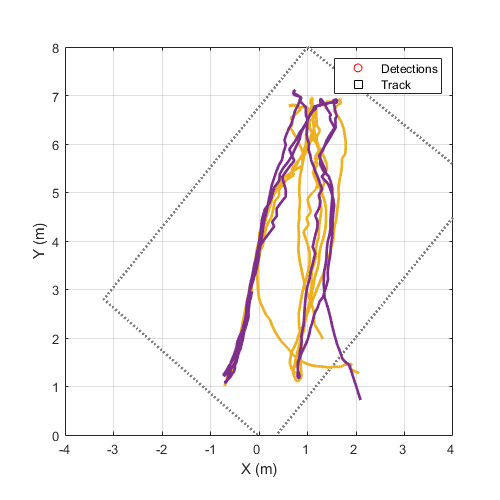}}
\subfloat[12 Channels]{\label{fig:FB_2T_12RX_RA}
\includegraphics[width=0.19\linewidth]{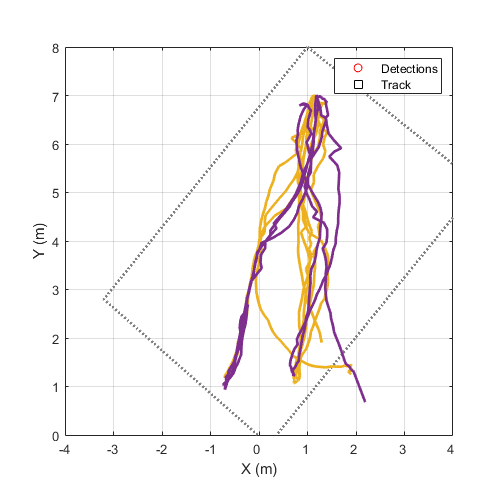}} 
\subfloat[8 Channels]{\label{fig:FB_2T_8RX_RA}
\includegraphics[width=0.19\linewidth]{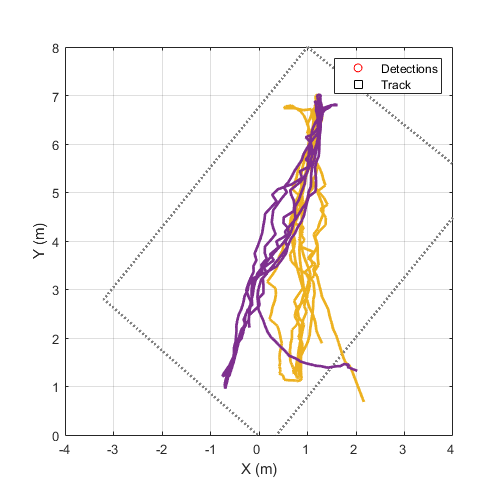}}
\subfloat[6 Channels]{\label{fig:FB_2T_6RX_RA}
\includegraphics[width=0.19\linewidth]{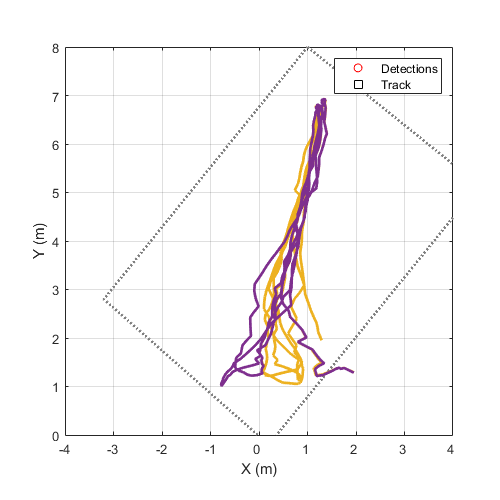}} 
\subfloat[4 Channels]{\label{fig:FB_2T_4RX_RA}
\includegraphics[width=0.19\linewidth]{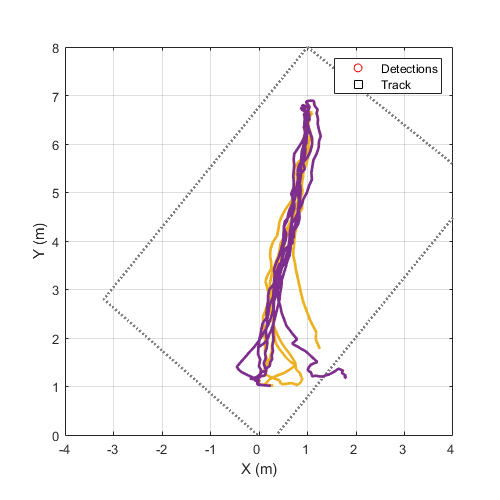}}
\caption{Number of channels analysis (i.e., angular resolution) for 2 persons; pipeline with detection from RA maps}
\label{fig:re2TFBRA_CHN}
\end{figure*}

\begin{figure*}[!t]
\centering
\subfloat[15 Channels]{\label{fig:FB_2T_15RX_RD}
\includegraphics[width=0.19\linewidth]{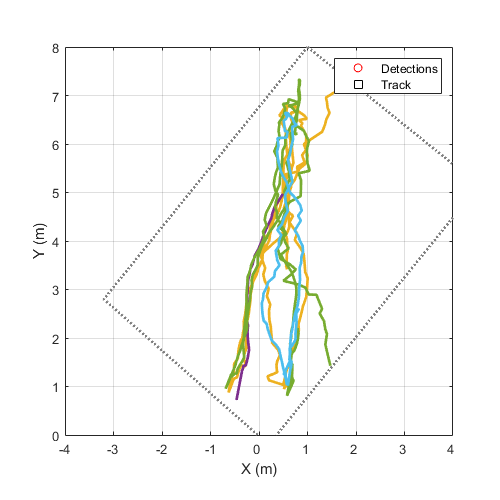}}
\subfloat[12 Channels]{\label{fig:FB_2T_12RX_RD}
\includegraphics[width=0.19\linewidth]{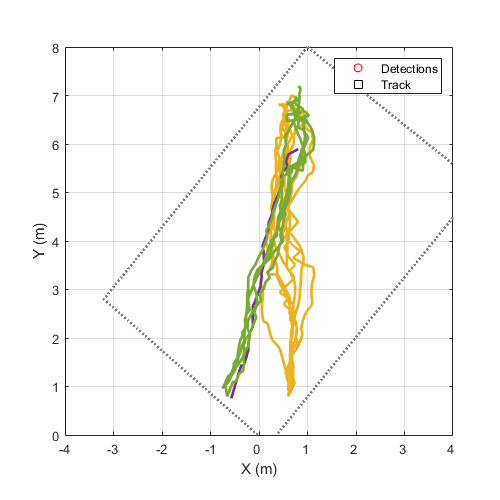}} 
\subfloat[8 Channels]{\label{fig:FB_2T_8RX_RD}
\includegraphics[width=0.19\linewidth]{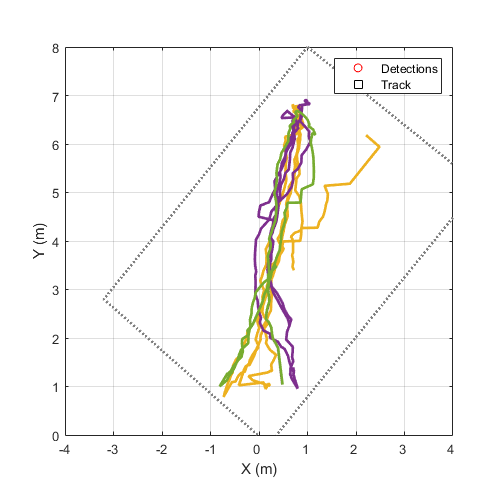}}
\subfloat[6 Channels]{\label{fig:FB_2T_6RX_RD}
\includegraphics[width=0.19\linewidth]{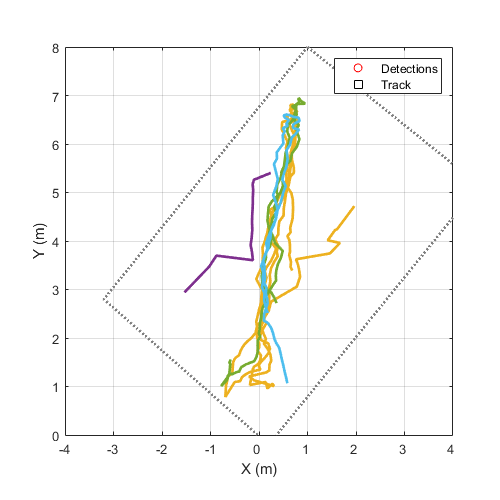}} 
\subfloat[4 Channels]{\label{fig:FB_2T_4RX_RD}
\includegraphics[width=0.19\linewidth]{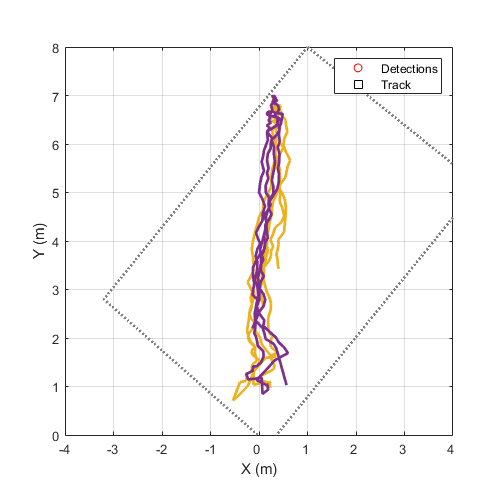}}
\caption{Number of channels analysis (i.e., angular resolution) for 2 persons; pipeline with detection from RD maps}
\label{fig:re2TFBRD_CHN}
\end{figure*}

However, the importance of angular resolution is very clear in multiple target scenarios. Tracking results for a different number of channels are evaluated for two people walking side by side away from and towards the radar. As shown in Fig. \ref{fig:re2TFBRA_CHN} \& Fig. \ref{fig:re2TFBRD_CHN}, the two tracks start to merge into one earlier, i.e., closer to the radar, when reducing the channels available for angular estimation in the RA-based pipeline. From Fig. \ref{fig:re2TFBRA_CHN}, the separation of two targets at a close distance needs at least 8 channels, as using 4 or 6 channels produces unrecognizable trajectories. As in Fig. \ref{fig:re2TFBRD_CHN}, the trajectories can be recognized as two targets when using 15 channels. However, the issue of tracking ID switch appears. The detection on RD maps relies more on high angular resolution when two targets are close together. 
When the spacing of the targets is less than the angular resolution in a farther range bin, the separation of the two detected clusters is still a challenge. In such cases, more advanced methods for the separation of each individual may be explored \cite{SuperimposedMaskguidedContrastivexu2023}, or alternative approaches for tracking multiple targets as a group \cite{GroupedPeopleCountingren2023}.

\subsection{Tracking results of selected scenarios}
\begin{figure}[h]
\centering
\subfloat[Pipeline with detection on RA maps]{\label{fig:2T_4mRA}\includegraphics[width=0.53\linewidth]{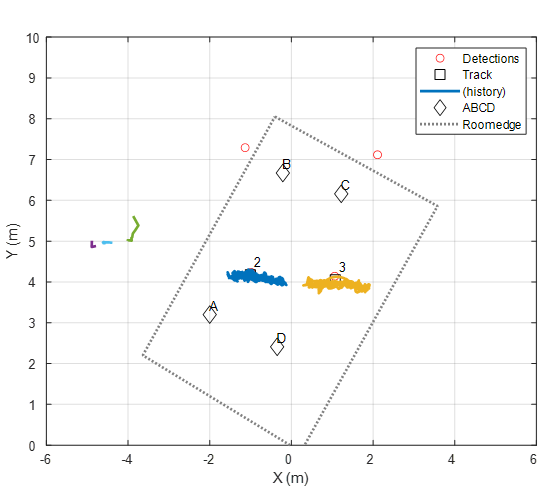}}

\subfloat[Pipeline with detection on RD maps]{\label{fig:2T_4mRD}\includegraphics[width=0.53\linewidth]{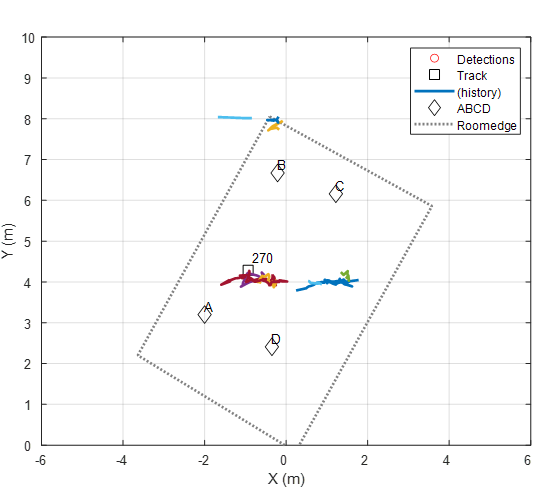}}
\caption{Tracking results for 2 persons walking in the tangential direction with respect to the radar line of sight}
    \label{fig:2T4m}
\end{figure}

In order to compare the two processing pipelines, an example of tracking results is presented for a specific scenario. As shown in Fig. \ref{fig:2T4m}, two targets start moving from the sides of the room and walk tangentially to the radar line of sight, approaching each other and then separating at about 4 m. The pipeline based on detection on RA maps shows complete trajectories when targets are getting close, at a distance of about 30 cm, when using 15 channels for angle estimation.
The root mean square error (RMSE) is 15.84 cm.
On the contrary, the pipeline based on detection on RD maps could not reconstruct the complete trajectories and RMSE here is not meaningful. As the targets are moving in tangential directions, their Doppler contributions are in general reduced and easily mixed, which makes their separation based on RD maps rather challenging.


\subsection{Discussion on algorithm limitation}
Based on the presented experiments and analysis, the RA-based pipeline in Fig. \ref{fig:RAdiagram} appears to be more robust than the alternative based on RD maps. However, for good performances, this also requires multiple channels for finer angular resolution and is affected by the resolution variation with radial distance and angular separation from the foresight. The detection on RD maps shows less favorable performance to detect multiple targets at a close range. However, the Doppler information only depends on targets' movements and is not related to their range \& angular position, potentially leading to more reliable performance within the area under test.

It is also important to consider the effect of the available resolution on the performance. Doppler resolution depends on the number of chirps used and on RD maps it can be useful to distinguish the contributions from different body parts for later classification or individual identification. On RA maps, the Doppler resolution will impact the estimation of radial velocity 
and alter the measurement noise in the Kalman filter within the tracking stage. 
Additionally, the availability of multiple snapshots (i.e., more chirps/frames) will contribute to angular estimation by separating the signal and noise components when decomposing the covariance matrix. 
In this context, iterative techniques to update the covariance matrix with fewer snapshots could be beneficial \cite{BeamformingFastScanningdash2023}.



\section{Conclusion} 
\label{sec:conclusion}

In this paper, two processing pipelines for multiple target tracking for indoor human monitoring are presented and compared based on experimental data collected with a 24 GHz MIMO FMCW radar.
The pipeline based on detections on RA maps appears to be more robust than the alternative based on detections on RD maps, where the latter is shown to underperform in tracking targets located at close range bins. Different CFAR methods for indoor human detection are compared, with OS-CFAR showing the best performance, and OSCA-CFAR providing a viable alternative when the computational complexity may become a limitation. Varying the number of channels used in azimuth estimation is also compared, showing that at least 8 channels are needed when 2 people are walking at a close distance. Future work will consider integrating more advanced tracking algorithms into the pipeline and information from the cadence velocity diagram (CVD) to better estimate scenarios with grouped people clustered \cite{GroupedPeopleCountingren2023}. 



\section*{Acknowledgment}

This research was in part financially supported by Huawei Sweden Gothenburg Research Center. The authors thank the Huawei team Zhong Chen, Yanming Wu and Jingjing Chen for the technical discussions, and are grateful to the volunteers who helped with the data collection.


\bibliographystyle{IEEEtran}
\bibliography{IEEEabrv, reference}


\end{document}